\documentclass[onecolumn,aps,prl]{revtex4}
\usepackage{graphicx}
[1999/03/29 PSNFSS v.7.2
Times font as default roman
: S Rahtz]

\begin{document}
\renewcommand{\thefootnote}{\fnsymbol{footnote}}
\sloppy
\newcommand{\rp}{\right)}
\newcommand{\lp}{\left(}
\newcommand \be  {\begin{equation}}
\newcommand \ba {\begin{eqnarray}}
\newcommand \ee  {\end{equation}}
\newcommand \ea {\end{eqnarray}}

\title{Detecting anchoring in financial markets} 
\thispagestyle{empty}

\author{J\o rgen Vitting Andersen}
\affiliation{
Institut Non Lin\'eaire de Nice
1361 route des Lucioles, Sophia Antipolis 
F06560 Valbonne (FRANCE)}

\email{vitting@unice.fr}

\date{\today}

\begin{abstract}
Anchoring is a term used in psychology to describe the common human 
tendency to rely too heavily (anchor) on one piece of information
when making decisions. A trading algorithm inspired by biological 
motors, introduced by L. Gil\cite{Gil}, is suggested as 
a testing ground for  anchoring 
in financial markets. An exact solution of the algorithm is presented 
for arbitrary price distributions. Furthermore the algorithm is  
extended to cover the case of a market neutral portfolio, revealing 
additional evidence that anchoring is involved in the decision making 
of market participants. The exposure of arbitrage possibilities 
created by anchoring gives yet another illustration on the 
difficulty proving market efficiency by only considering   
lower order correlations in past price time series.
\end{abstract}

\maketitle

\vspace{1cm}

Behavioral finance has become a growingly influential subbranch 
of finance, stressing how human and social emotional biases 
can affect market prices. The field adapts a more pragmatic 
and complex view 
on financial markets, in contrast to standard finance 
theory where people 
rationally and independently on basis of full information 
try to maximize utility. The more realistic view comes at a 
cost however, since a multifactorial world is captured mainly via 
observations or postulates about human behavior. 

One of the first observations of anchoring was reported in the now classical 
experiment by Tversky and Kahneman\cite{Tversky}.  
Two groups of test persons were shown to give different mean estimates 
of the percentage of African nations in the United Nations, depending  
on the specific anchor of percentage suggested by the experimenters 
to the two groups.  
Evidence for human anchoring has since been reported 
in many completely different domains 
such as e.g. customer inertia in brand switching\cite{Ye} (old brand price act  
as an anchor), whereas other evidence 
come from studies 
on on-line auctions\cite{Dodonova} (people bid 
more for an item the higher the ``buy-now'' price) and anchoring in real 
estate prices\cite{Northcraft} (subjects appraisal values depend 
on arbitrary posted listing price of the house). In the 
context of financial markets anchoring has been observed via 
the so called ``disposition effect''\cite{Shefrin},\cite{Heilmann} which is 
the tendency for people to sell assets that have gained value 
and keep assets that have lost value. 
As noted in \cite{Weber} 
conclusive tests using real market data is usually difficult 
because the investors' expectations, as well as individual 
decisions, can not be controlled or easily observed. In 
experimental security trading however subjects were observed 
to sell winners and keep losers\cite{Weber}.

In order to get a more firm understanding of how 
aggregation of individual 
 behavior can give rise to measurable effects in a population in 
general and financial markets in particular, 
it would be interesting to model specific human traits on a micro 
scale and study the emergence of a dynamics with observable or  
even predictable effects on a macro scale. The hope would be 
to reproduce in models many of the mechanisms reported at work 
in behavioral finance. 
One step in this direction was done in \cite{Andersen1} where it 
was shown how consensus (called ``decoupling'' in \cite{Andersen1}) 
and thereby predictability could emerge due to mutual 
influence of the price in a 
commonly  traded asset, among a group of agents who
had initially different 
opinions.
Here another method is 
suggested which rigourously test for a different human trait 
introduced by behavioral finance, namely anchoring. The 
algorithm used was introduced by L. Gil \cite{Gil}, and is inspired 
from the way  
biological motors work by exploiting favourable brownian fluctuations to 
generate directed forces and move. Similar ideas were also introduced 
in \cite{Sornette} where it was shown how increments of uncorrelated 
time series can be predicted with a universal 75\% probability of 
success.

Specifically, assume an agent at every time step $t$ 
uses a fixed amount of his wealth to hold a long position 
in one out of $N$ assets. For simplicity $N=2$ will be used 
in the following, but the arguments can be extended to arbitrary 
$N$.  
Assume furthermore that the 
probability distribution functions (pdf's) of the {\em price} 
of the two assets, $P_1(A_1),P_2(A_2)$ are 
 stationary distributions. Instead of the usual assumption of a random 
walk of the returns, short time anchoring of prices at quasi 
static price levels is imposed.
No specific shape is assumed and the 
assets can be correlated or not, but any correlation is  
irrelevant for the following arguments. As noted in \cite{Gil} 
 the assumption of short term 
stationarity of prices can arise because of price 
reversal dynamics caused e.g. by monetary policies. As 
will be argued and tested for in the following , short term 
``stationarity'' in prices can also be created due to short 
term human memory as to when an asset is ``cheap'' or ``expensive''.  

Consider any given instantaneous fluctuation of the 
prices $(A_1,A_2)$ around their quasi static price 
levels 
$(\bar{A_1},\bar{A_2})$.  
Classifying the $2^N$ different cases according 
whether $A_i < \bar{A_i}$ or $A_i > \bar{A_i}$, one 
has for $N=2$ the four different configurations $x_i$:

\begin{center}
\begin{tabular}{|c|c|c|c|}
\hline
\ \ \ \ \  $x_1$ \ \ \ \ & \ \ \ \ \ $x_2$ \ \ \ & 
\ \ \ \ \  $x_3$ \ \ \ \ & \ \ \ \ \ $x_4$ \ \ \ \\\
 &  & & \\
   $A_1$ --- \ \  \ \ &   $A_1$ --- \ \ \ \ & 
  \ \  \ \ &   \ \ \ \ \\
 \  \ \ $\updownarrow  dA_1$   \ \ &   \ \ \ $\updownarrow dA_1$ \ \ &
 \  \ \    \ \ &   \ \ \  \ \ \\
$\bar{A_1}$ \ \--------  \ \ &  $\bar{A_1}$ \ \-------- &
$\bar{A_1}$ \ \--------  \ \ &  $\bar{A_1}$ \ \-------- \\
 \  \ \    \ \ &   \ \ \  \ \ &
 \  \ \ $\updownarrow  dA_1$   \ \ &   \ \ \ $\updownarrow dA_1$ \ \ \\
   \ \  \ \ &   \ \ \ \ &
 $A_1$ --- \ \  \ \ &   $A_1$ --- \ \ \ \ \\
 &  & & \\
   \ \  \ \ &   $A_2$ --- \ \ \ \ &
   \ \  \ \ &   $A_2$ --- \ \ \ \ \\
 \  \ \    \ \ &   \ \ \ $\updownarrow dA_2$ \ \ &
 \  \ \    \ \ &   \ \ \ $\updownarrow dA_2$ \ \ \\
$\bar{A_2}$ \ \--------  \ \ &  $\bar{A_2}$ \ \-------- &
$\bar{A_2}$ \ \--------  \ \ &  $\bar{A_2}$ \ \-------- \\
 \  \ \ $\updownarrow  dA_2$   \ \ &   \ \ \  \ \ &
 \  \ \ $\updownarrow  dA_2$   \ \ &   \ \ \  \ \ \\
   $A_2$ --- \ \  \ \ &    \ \ \ \ &
   $A_2$ --- \ \  \ \ &    \ \ \ \ \\
 &  & & \\
\hline
\end{tabular}
\end{center}

In steady state the probability flux into a given 
configuration $x_i$ equals the probability flux out of 
that configuration:
\ba
\sum_j P(x_j) P(x_j \rightarrow x_i) & = & 
\sum_j P(x_i) P(x_i \rightarrow x_j) \\
\label{master_eq}
\ea
The averaged return per time unit in the steady state, $R_{\rm av}$, 
is then given by 
\ba
R_{\rm av} & = & \sum_{i=1}^4 \sum_{j=1}^4 P(x_i) P(x_i \rightarrow x_j) 
r_{\rm av}(x_i \rightarrow x_j) 
\label{Rav_per_time}
\ea
with 
$r_{\rm av}(x_i \rightarrow x_j)$ the averaged return gained/lost 
in the transition $x_i \rightarrow x_j$. For each configuration 
$x_i$ one is assumed to hold a long position of {\em either} asset 1 
{\em or} 
asset 2. Let $s=i$ be a state 
variable indicating that one is long one position of asset $i$. 
Then 
\ba
r_{\rm av}(x_i \rightarrow x_j) 
 & = & P(s=1|x_i) 
r_{\rm av}(x_i \rightarrow x_j|s=1) 
 + P(s=2|x_i) 
r_{\rm av}(x_i \rightarrow x_j|s=2) 
\label{rav_transition}
\ea
where 
$P(s=i|x_j)$ 
denotes the probability holding asset $i$ 
given the knowledge to be in configuration $x_j$.  
$r_{\rm av}(x_i \rightarrow x_j|s=k)$  
denotes
the averaged return in steady state holding asset $k$ with 
a transition from configuration $x_i$ to $x_j$ and is given 
by:  
\ba
r_{\rm av}(x_j \rightarrow x_k|s=k) 
 & = & \int dA_k \int dA_k^{'}
\ln  ( {A_k^{'} \over A_k} )  P(A_k^{'}|x_i) 
  P(A_k|x_j) 
\label{rav_transition_asset_i}
\ea
$P(A_k|x_i)$ denotes the probability to get the price $A_k$ 
{\em conditioned} on being in configuration $x_i$. For example
knowing to be in configuration $x_1$ one has: 
$
  P(A_2|x_1) 
 =  
{  P(A_2) \theta (A_2 \leq 0) \over 
\int_{- \infty}^0  P(A_2^{'})  dA_2{'} }
$
with 
$\theta (A_2 \leq 0)$ a Heaviside function. 
Using (\ref{Rav_per_time}-\ref{rav_transition_asset_i}) 
the general expression for the average return gained by
the algorithm takes the form:
\ba
R_{\rm av} & = & 
\sum_{i=1}^4 \sum_{j=1}^4 \sum_{s=1}^2 P(x_i) P(x_i \rightarrow x_j) 
  P(s|x_i) 
 \int dA_k \int dA_l
\ln  ( {A_l \over A_k} )  P(A_l|x_i) 
  P(A_k|x_j) 
\label{Rav_per_time_final_expression}
\ea

The corresponding risk measured by the averaged 
standard deviation of the return is given by: 
\ba
\sigma^2 & = & <(r-R_{\rm av})^2> \\
   & = & 
\sum_{i=1}^4 \sum_{j=1}^4 \sum_{s=1}^2 P(x_i) P(x_i \rightarrow x_j) 
  P(s|x_i) 
 \int dA_k \int dA_l
\ln  ( {A_l \over A_k} )^2  P(A_l|x_i) 
  P(A_k|x_j) - R_{\rm av}^2 
\label{sigmaav_per_time_final_expression}
\ea

The ``trick'' of the algorithm consists in breaking 
the symmetry by always choosing $P(s|x_i)$ according to 
the following rules:
\ba
P(s=1|x_1)=0 ; \ \ p(s=2|x_1)=1 ; 
& & \ \ p(s=1|x_2)= p(s=2|x_2) = 1/2 ; \nonumber \\ 
P(s=1|x_3) = p(s=2|x_3)= 1/2 ; 
& &  \ \ p(s=1|x_4)=1; \ \  p(s=2|x_4) = 0   
\label{symmetry_breaking_eqs}
\ea
That is, if not already long, one always take a long position of 
asset 2 (1) whenever configuration $x_1$ ($x_4$) happens, since the 
asset is undervalued in this case. Likewise if one is long of asset 1 (2) 
whenever configuration $x_1$ ($x_4$) happens one sell that asset 
since it is overvalued.  
To illustrate the algorithm consider the simplest case where $P(A_i)$ 
take only two values $\bar{A_i} \pm dA_i$ with equal probability 
$1/2$.
Inserting 
\ba
P(A_1|x_1)=\delta (\bar{A}_1+ dA_1); \ \
P(A_1|x_2)=\delta (\bar{A}_1+ dA_1); \ \
P(A_1|x_3)=\delta (\bar{A}_1- dA_1); \ \
P(A_1|x_4)=\delta (\bar{A}_1- dA_1); \nonumber \\
P(A_2|x_1)=\delta (\bar{A}_2- dA_2); \ \
P(A_2|x_2)=\delta (\bar{A}_2+ dA_2); \ \
P(A_2|x_3)=\delta (\bar{A}_2- dA_2); \ \
P(A_2|x_4)=\delta (\bar{A}_2+ dA_2)  
\label{P_bin_eqs}
\ea
and $P(x_i) = P(x_i \rightarrow | x_j)= 1/4$ into
(\ref{Rav_per_time_final_expression}) one gets the  
averaged return:
\ba
R_{\rm av}^{\bar{A_i} \pm dA_i} & = & 
1/8 \ \  [ 
\ln  ( {\bar{A}_1 + dA_1 \over \bar{A}_1 - dA_1 } ) 
+ \ln  ( {\bar{A}_2 + dA_2 \over \bar{A}_2 - dA_2 } ) 
 ] 
\label{Rav_binomial}
\ea
 with a variance given by
\ba
(\sigma_{\rm av}^{\bar{A_i} \pm dA_i})^
2 & = & 
15/64 \ln^2  ( {\bar{A}_1 + dA_1 \over \bar{A}_1 - dA_1 } ) 
+ 15/64 \ln^2  ( {\bar{A}_2 + dA_2 \over \bar{A}_2 - dA_2 } ) 
- 1/32 \ln  ( {\bar{A}_2 + dA_2 \over \bar{A}_2 - dA_2 }  
 {\bar{A}_1 + dA_1 \over \bar{A}_1 - dA_1 } ) 
\label{var_av_binomial}
\ea

\begin{figure}
\includegraphics[width=8cm]{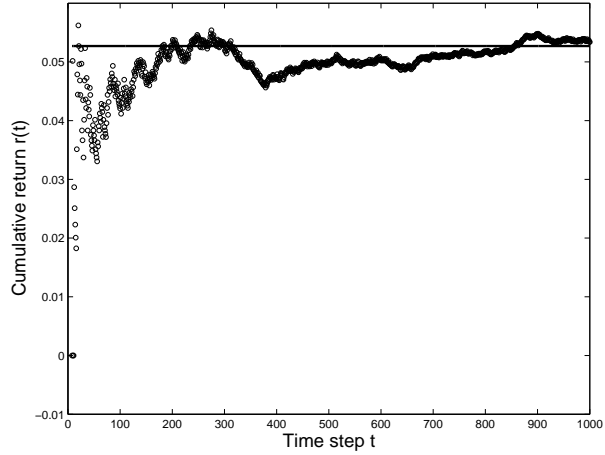}
\caption{\protect\label{Fig1}
Averaged return $r(t)$ as a function of time $t$. 
Circles represent the result obtained by using 
the algorithm (\ref{symmetry_breaking_eqs}) with 
$(\bar{A_1}=\bar{A_2}=1.0, dA_1=dA_2= 0.11)$, 
and memory $m=5$. Solid line represents the analytical 
expression (\ref{Rav_binomial}).  
}
\end{figure}

In order to check the algorithm 
(\ref{symmetry_breaking_eqs}) with the expressions 
(\ref{Rav_binomial}), (\ref{var_av_binomial}) 
random price time series
$P(A_i) = \bar{A_i} \pm dA_i$
(with the randomness 
stemming from the sign of $dA_i$) 
were generated with fixed values  
of 
$\bar{A_i},dA_i$. 
Figure~\ref{Fig1} 
 shows the average return obtained using the algorithm as 
a function of time. 
In order to make the classification as indicated in table~1,  
the averaged value of $\bar{A_i}$ 
was estimated as in \cite{Gil} using an average over the last 
$m$ price values. 
As seen after a transient the averaged 
return reaches the steady state expression 
(\ref{Rav_binomial}) as it should.

\begin{figure}
\includegraphics[width=8cm]{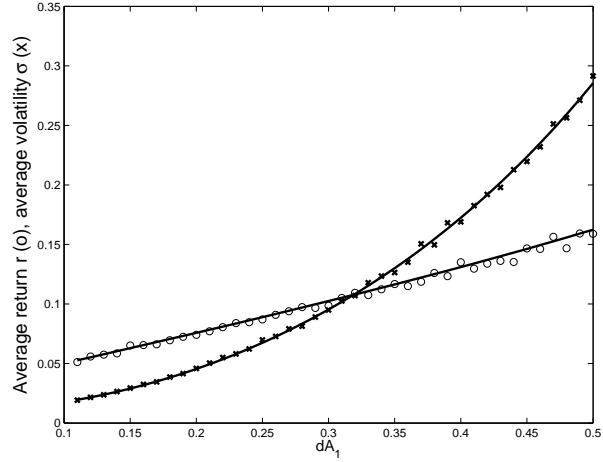}
\caption{\protect\label{Fig1inset}
Averaged return 
$R_{\rm av}^{\bar{A_i} \pm dA_i}$ (circles) and 
volatility 
$(\sigma_{\rm av}^{\bar{A_i} \pm dA_i})^2$ (crosses) versus 
$dA_1$ 
The data points were obtained in steady state  
 by the algorithm 
(\ref{symmetry_breaking_eqs}) 
using a memory $m=5$ to 
estimate $\bar{A_i}$ from which in turn the classification 
was made according to figure~1. 
The random price time series (with the randomness 
stemming from the sign of $dA_i$) were generated with fixed values 
$(\bar{A_1}=\bar{A_2}=1, dA_2=0.11)$. Solid lines 
represent analytical results (\ref{Rav_binomial}) ,
(\ref{var_av_binomial})
.
}
\end{figure}

The points in figure~\ref{Fig1inset} 
represent the steady state results obtained by the algorithm for  
the averaged return  
$R_{\rm av}^{\bar{A_i} \pm A_i}$ 
and 
volatility 
$(\sigma_{\rm av}^{\bar{A_i} \pm A_i})^2$ 
 versus $dA_1$. As seen the simulation results of the algorithm 
agree with the expressions
(\ref{Rav_binomial}) and 
(\ref{var_av_binomial}) represented by solid lines.

\begin{figure}
\includegraphics[width=8cm]{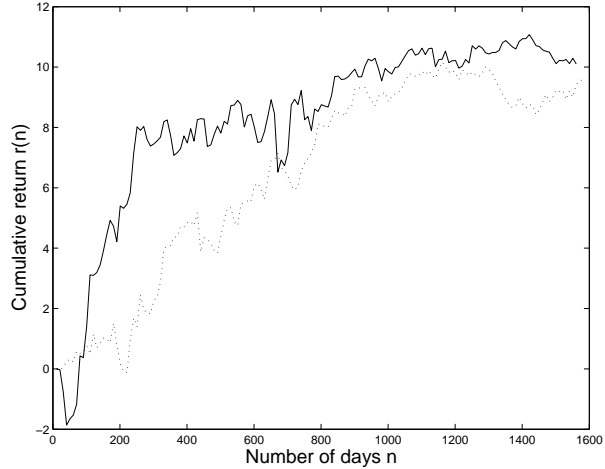}
\caption{\protect\label{Fig2}
Cumulative return of the market neutral algorithm applied to 
daily price 
data of the Dow Jones stock index (dotted line), as well as the CAC40 stock 
index (solid line) over the period 3/01/2000-2/5/2006. 
First half of the time period for the Dow Jones index 
was used in sample to determine the best choice 
among three values of the parameter 
$m=5,10,15$ days. Second half of the time period for the 
Dow Jones index as well as full period of the CAC40 index was 
done out of sample with $m=10$. 
As a measure of the performance of the algorithm,  
the Sharpe ratio was found to be 2.0 and 2.94 for the Dow Jones, 
respectively CAC40 price time series. A trading cost of 0.1\% 
was included for each transaction.
}
\end{figure}

The algorithm was then applied to real market data. 
However as  
noted in \cite{Gil} a general  
problem arises because of long term  
drifts in anchor of the price, $\bar{A_i}$,  
which is never truely ``quasi-static''.  
I.e. $\bar{A_i}$ is time 
dependent, and for sufficient strong drifts 
the return of the algorithm was 
then shown to vanish. In order to circumvent 
this obstacle the algorithm was modified so as always to 
be market neutral {\em independent} of  
any drift 
 the portfolio of the $N$ assets might perform. Figure~\ref{Fig2}
shows the market neutral algorithm applied to real market 
data of the Dow Jones stock index, as well as the CAC40 stock 
index. First half of the time period for the Dow Jones index 
was used in sample to determine the best choice 
among three values of the parameter 
$m=5,10,15$ days. Only three possible values corresponding
to one, two or three weeks were probed in sample. Since 
the present paper look into any possible impact coming 
from human anchoring, using only daily or 
 weekly data seems a priori 
justified since the higher the 
frequency of the trading (say seconds/minutes) 
the more computer dominated becomes the trading. 
In order to look for arbitrage 
possibilities weekly data was used so as to 
avoid impact of transaction costs by trading too often.  
Another reason only to have looked at weekly data is 
 because of the main claim put forward in this paper where  
 market participants by actively following the 
price thereby create a subjective reference (anchor) 
and memory of  
 when an asset is ``cheap'' or ``expensive''. 
Several studies on the persistence of human memory 
have reported sleep as well as post-training wakefulness 
before sleep, to play an important role in the 
offline processing and consolidation of memory\cite{Peigneux}. 
It therefore makes sense to think that conscious as well 
as unconscious mental processes influence 
the judgements of people who 
specializes in active trading on a day-to-day basis.
The out of sample profit from the market neutral 
trading algorithm (with transaction costs taking 
into account) on the CAC40 index as well 
as the second period performance on the Dow Jones index, 
gives evidence that anchoring does indeed play a dominant role on 
the weekly price fixing of the Dow Jones and CAC40 stock 
markets, 
and reconfirms the claim in \cite{Gil} where 
especially the policy imposed by the European Monetary 
System was shown to lead to arbitrage possibilities. 
The results also 
 gives yet another illustration on the 
difficulty proving market efficiency by only considering   
lower order correlations in past price time series.

In conclusion 
a trading algorithm inspired by biological 
motors and introduced by L. Gil, is suggested as 
a testing ground for  anchoring 
in financial markets. An exact solution of the algorithm was 
found  
for arbitrary price distributions and the algorithm was  
extended to cover the case of a market neutral portfolio. 
The exposure of arbitrage possibilities by the market neutral algorithm 
reveals
additional evidence that anchoring is indeed involved in the decision making 
of market participants. 

The author is grateful to L. Gil for valuable discussions.

{}


\vskip -0.7cm

\begin{thebibliography}{}
\bibitem{Gil}L. Gil, arXiv:0705.2097 {2007}

\bibitem{Tversky} A. Tversky and D. Kahneman, 
Science {\bf 185}, 1124 (1974).

\bibitem{Ye} G. Ye, 
SSRN-id548862, (2004).

\bibitem{Dodonova}
A. Dodonova and Y. Khoroshilov,
Applied Economics Letters {\bf 11}, 307 (2004).

\bibitem{Northcraft}
G. B. Northcraft and M. A. Neale,
Organizational Behavior and Human Decision Processes, {\bf 39}, 84 (1987).

\bibitem{Bofinger}
P. Bofinger and R. Schmidt, 
Discussion Paper Series - Centre For Economic Policy Research London, 
ISSU 4235 (2004).

\bibitem{Shefrin}
H. M. Shefrin and M. Statman 
Journal of Finance {\bf 40}, 777 (1985).

\bibitem{Heilmann}
K. Heilmann, V. Laeger and A. Oehler,
Proceedings of the 25th Annual Colloquium, IAREP, Wien (2000).

\bibitem{Weber}
M. Weber and C. F. Camerer,
Journal of Economic Behavior \& Organization {\bf 33}, 167 (1998).

\bibitem{Andersen1}
J. V. Andersen and D. Sornette,
Europhys. Lett. {\bf 70}, 697 (2005).

\bibitem{Sornette}
D. Sornette and J. V. Andersen,
Int. J. Mod. Phys. {\bf C11}, 713 (2000).

\bibitem{Peigneux}
P. Peigneux, P. Orban, E. Balteau, C. Degueldre and A. Luxen,
PLoS Biology, Vol. 4, No.4, e100 doi:10.1371/journal.pbio.0040100 
(2006).

\end{thebibliography}
\end{document}